\documentclass[twocolumn,prl]{revtex4}
\usepackage{bm}
\usepackage{graphicx}
\begin{document}
\title{On the Spin-Hall effect in a dirty Rashba semiconductor}
\author{O.Bleibaum}
\email{olaf.bleibaum@physik.uni-magdeburg.de}
\affiliation{Institut f\"ur Theoretische Physik, 
Otto-von-Guericke Universit\"at, 39016 Magdeburg, Germany}
\begin{abstract}
The impact of the boundary condition on the solution
of the spin diffusion equations in a dirty Rashba semiconductor
is reinvestigated. To this end first diffusion equations, which take into 
account the coupling between spin and charge, are derived. 
The investigation of the solution to the spin-diffusion equation
near the boundary shows that the spin-current tensor vanishes 
discontinuously at the boundary. As a consequence, there is no
spin-Hall effect in a dirty Rashba semiconductor.
\end{abstract}
\maketitle
Spintronics is a rapidly developing field, which has the objective to learn 
more about the interaction of the electron spin with its solid state 
environment. It is stimulated by the fact that the electron spin can also
be manipulated electrically and thus be utilized in electrical devices. The 
interaction, which provides access to electric spin manipulation is the 
spin-orbit interaction, which is realized by the Rashba interaction in the 
simplest situation\cite{Zutic}.

The injection of spins into nonmagnetic semiconductors, however, has proven 
to be a difficult task. Therefore, much attention is paid to the investigation 
of effects, which permit the electrical creation of non-vanishing magnetizations
in the sample. Already the first investigations on the impact of a lateral
electric field on a two-dimensional Rashba semiconductor have shown that such
effects in principle exist\cite{Edelstein}. According to 
Ref.[\onlinecite{Edelstein}], e.g., the lateral field leads to a homogeneous 
magnetization in the two-dimensional plane. Similar results have been obtained 
in the Refs.[\onlinecite{Inoue}] and [\onlinecite{Halperin}] recently.

An effect, which might be useful in creating non-vanishing magnetizations, is the spin-Hall effect. The latter manifests itself in a non-vanishing 
magnetization at the sample boundaries in the presence of an electric field.
In that it differs from the spin accumulation, which is present in the bulk. 
Whereas there seems to be some agreement over the magnitude of the spin
accumulation in a Rashba semiconductor, the spin-Hall effect is still a 
rather controversial issue. On the one hand an intrinsic spin-Hall effect has 
been suggested in the Refs.[\onlinecite{Sinova}] and [\onlinecite{Murakami}]
and investigated further in a number of papers. These authors argue, that the
expection value of the spin-current operator, defined by the equation
\begin{equation}\label{I1}
{\hat J}_{ik}=\frac{1}{2}\{{\hat v}_i,\sigma_k\},
\end{equation}
is non-vanishing in a clean Rashba semiconductor in the presence of an 
electric field $F$, and that therefore a magnetization is created at the 
boundaries. (${\hat v}_i$ is the velocity operator, $\sigma_k$ are the Pauli 
matrices, $i=1,2$, $k=1,2,3$. The indices 1 and 2 and the labels x and y are 
used simultaneously to characterize the vectors in the 2-d plane, the index 
3 and the label z to characterize the perpendicular direction). The spin-Hall 
conductivity, defined by the ratio $\sigma_H=J_{yz}/F$, is universal in clean 
systems, that is independent of the magnitude of the Rashba interaction.
In dirty systems, in which the Rashba level splitting is small compared to the
disorder energy, the spin-Hall conductivity vanishes\cite{Loss,Bauer,Burkov}.
Thus, it is expected that there is no spin-Hall effect in such systems.
On the other hand spin-Hall effect has been discussed in Ref.
[\onlinecite{Damker}] in the hopping regime. Doing so, it has been assumed
that the Rashba level splitting can be ignored. Thus, the system of Ref.
[\onlinecite{Damker}] is always in the dirty limit.

When investigating the differences between the approaches of the Refs.
[\onlinecite{Sinova}]-[\onlinecite{Burkov}] and the approach of Ref.
[\onlinecite{Damker}] we note the following: whereas a diffusion equation 
for the spin density has been derived in Ref.[\onlinecite{Damker}] the
conclusions of the Refs.[\onlinecite{Sinova}]-[\onlinecite{Burkov}]
are based only on the investigation of the spin-current tensor (\ref{I1}).
However, it is unclear how the tensor (\ref{I1}) is related to observable
quantities\cite{Rashba}. Therefore, there are no predictions of the magnitude 
of the magnetization at the sample boundaries. Moreover, the expection value 
of the spin-current is non-vanishing even in equilibrium\cite{Rashba}. This
fact raises the question whether the spin-current tensor is well defined at all
\cite{Rashba}. In Ref.[\onlinecite{Damker}], by contrast, the spin-current
tensor has been read off from the diffusion equation. To this end, the
authors of Ref.[\onlinecite{Damker}] have taken the point of view, that
all terms, which can be cast into the form of a derivative, contribute
to the divergence of the spin-current tensor. Using this approach they
calculate the solution to the spin-diffusion equation and make detailed
predictions to the magnitude of the magnetization at the sample boundaries.  
However, also this approach is not unproblematic. Since the 
diffusion equations yields only the divergence of the spin-current tensor,
the tensor read off from the diffusion equation is only determined up to a
curl and boundary currents had to be assumed to be non-existent at all. 
Moreover, 
the spin accumulation has entirely been ignored.

Since there seems to be no rigorous  method, which could be used to define the 
spin-current tensor unambiguously, there seems to be no other way than to 
investigate the consequences of the definition (\ref{I1}) and  to compare 
them with experiments. Acording to Ref.[\onlinecite{Damker}] 
the spin-current tensor determines the boundary conditions in the
diffusive regime. Thus, it is  natural to focus on boundary effects. 
To get further insight into the implications of the definition
(\ref{I1}) and into the question, in which way the spin current tensor 
determines  the boundary conditions, we investigate the magnetization in 
a dirty Rashba semiconductor in a half plane. To this
end we derive spin-diffusion equations for a two dimensional system
and solve these equations subjected to the boundary condition that the 
spin-current tensor vanishes at the boundary. 

Diffusion equations, which
also take into account the coupling between spins and charges, have been 
first derived in the Refs.[\onlinecite{Halperin}] and [\onlinecite{Burkov}].
Here, we use the same method as in Ref.[\onlinecite{Burkov}]. The 
Hamiltonian is given by the equation
\begin{equation}\label{I2}
H=\frac{{\bm{\hat p}}^2}{2m}-({\bm{\sigma}},{\bm N}\times{\bm{\hat p}})+
V({\bm x}).
\end{equation}
Here ${\bm{\hat{p}}}$ is the momentum operator, ${\bm N}=N{\bm e}_z$, and
$m$ is the effective mass. $V({\bm x})$ is a random potential with Gaussian 
distribution function, zero mean and standard deviation
\begin{equation}\label{I3}
\langle\langle V({\bm x})V({\bm x'})\rangle\rangle=\frac{\hbar}{2\pi\nu\tau}
\delta({\bm x}-{\bm x'}).
\end{equation}
$\nu$ is the density of states per spin and $\tau$ the single particle
relaxation time. The propagation of single particle excitations is described 
by the retarded and advanced Green's functions. These functions are the 
solution to the equation 
\begin{equation}\label{I4} 
((\pm i\frac{\hbar s}{2}+E){\bm 1}-H)G^{R/A}({\bm x}, {\bm x'}|E,s)={\bm 1}
\delta({\bm x}-{\bm x'}).
\end{equation}
Here ${\bm 1}$ is the $2\times 2$ unit matrix, $s$ is a fixed frequency, 
and $E$ is the total particle energy. For the calculation of their 
configuration averages we use the self-consistent Born approximation. In 
this approximation the Fourier-transform $g^{R}({\bm k}|E,s)$ of the 
configuration averaged retarded Green's functions 
${\bar G}^{R}({\bm x}-{\bm x'}|E,s)$ is given by the equation
\begin{eqnarray}\label{I5}
g^R({\bm k}|E,s)&=&g_+({\bm k}|E,s)g_-({\bm k}|E,s)\nonumber\\
& &\times (g^{-1}({\bm k}|E,s){\bm 1}-({\bm\sigma},{\bm N}
\times{\bm p})),
\end{eqnarray}
where $g_\pm^{-1}({\bm k}|E,s)=i\hbar s/2+E-E_{\pm}(k)+i\hbar/2\tau$,
$E_{\pm}(k)=\hbar^2 k^2/2m\pm|N|\hbar k$ and 
$g({\bm k}|E,s)=g_+({\bm k}|E,s)|_{N=0}$. The Fourier transform 
$g^A({\bm k}|E,s)$ of the configuration averaged advanced Green's function
${\bar G}^A({\bm x}-{\bm x'}|E,s)$ is obtained from Eq.(\ref{I3}) by hermitian
conjugation.

The evolution of the particle density and the spin density is described by a generalized diffusion equation. The densities are defined by the 
relationship
\begin{equation}\label{I6}
f_{\alpha\alpha'}({\bm x}|E,t)=\mbox{tr}(\psi^+_{\alpha}({\bm x},t)\psi_{\alpha'}({\bm x},t)\rho_0(E)),
\end{equation}
where $\psi^+_{\alpha}({\bm x},t)$ and $\psi_{\alpha}({\bm x},t)$ are creation and annihilation operators for particles with spin $\alpha$, and $\rho_0(E)$ is the initial density matrix. The particle density is given by the equation
$n({\bm x}|E,t)=\mbox{tr}f({\bm x}|E,t)$, the spin density by the relationship
$S_i({\bm x}|E,t)=\mbox{tr}(\sigma_if({\bm x}|E,t))$. Note, that this 
definition differs from the conventional definition by a factor $2/\hbar$.
The equation of motion for the quantity $f$ takes the form
\begin{equation}\label{I7}
\int dx_1\Gamma^{\alpha\alpha_1}_{\alpha'\alpha_2}({\bm x},{\bm x_1}|E,s)
f_{\alpha_1\alpha_2}({\bm x_1}|E,s)=f^0_{\alpha\alpha'}({\bm x}|E)
\end{equation}
after Laplace transformation with respect to times ($t\to s$). 
A summation with respect to double indices has to be performed.
The quantity $f^0_{\alpha\alpha'}({\bm x}|E)$ is the initial condition.
The kernel $\Gamma$ in Eq.(\ref{I7}) is given by the equation
\begin{eqnarray}\label{I8}
\tau\Gamma_{\alpha'\alpha_2}^{\alpha\alpha_1}\hspace{-2ex}&(&
\hspace{-2ex}{\bm x},{\bm x'}|E,s)
=\delta({\bm x}-{\bm x'})\delta_{\alpha\alpha_1}
\delta_{\alpha'\alpha_2}\nonumber\\\
&-&\hspace{-1ex}
\frac{\hbar}{2\pi\nu\tau}{\bar G}^R_{\alpha\alpha_1}({\bm x}-{\bm x'}|E,s)
{\bar G}^A_{\alpha_2\alpha'}({\bm x'}-{\bm x}|E,s)
\end{eqnarray}
in the ladder approximation.
To simplify Eq.(\ref{I7}) we use the hydrodynamics expansion. Doing so, we 
obtain the following diffusion equations after an inverse Laplace 
transformation
%
\begin{equation}\label{I9}
\frac{dn}{dt}-D\Delta n+\Omega\tau({\bm N}\times{\nabla},{\bm S})=0
\end{equation}
\begin{eqnarray}\label{I10}
\frac{d{\bm S}}{dt}+{\bm \Omega}\cdot{\bm S}&-&D\Delta{\bm S}-{\omega_s}
({\bm N}\times{\bm\nabla})\times{\bm S}\nonumber\\
&+&\Omega\tau{\bm N}\times{\bm{\nabla}} n=0.
\end{eqnarray}
%
These equations agree with those of Ref.[\onlinecite{Burkov}]. 
Here $D=E\tau/m$ is the diffusion
coefficient, $\omega_s=4mD/\hbar$ and ${\bm\Omega}_{ik}=\Omega\delta_{ik}(1+
\delta_{i3})$, where $\Omega=4m^2N^2D/\hbar^2$.

To calculate the spin-current density we
use the expression
\begin{eqnarray}\label{I11}
J_{ik}({\bm x}|E,s)&=&\frac{\hbar}{2\pi\nu\tau}\mbox{tr}
\int d{\bm y} {\bar G}^A({\bm y}-{\bm x}|E,0)
{{\hat J}_{ik}}\nonumber\\
& &\hspace{1em}\times{\bar G}^R({\bm x}-{\bm y}|E,0)f({\bm y}|E,s),
\end{eqnarray}
which is valid in the ladder approximation, provided we are
interested in time scales which are large compared to $\tau$\cite{Halperin}.
To simplify the integral we use the same hydrodynamic expansion as 
before. Doing so, we obtain the following expressions for components
of the spin-current tensor
\begin{equation}\label{I12}J_{xx}=-D\nabla_x S_x-
\frac{N\omega_s}{2}S_z,
\end{equation}
\begin{equation}\label{I13}
J_{xy}=-D\nabla_xS_y+\Omega\tau N n,
\end{equation}
\begin{equation}\label{I14}
J_{xz}=-D\nabla_xS_z+\frac{N\omega_s}{2}S_x-\frac{\omega_s\tau N^2}{2}
\nabla_y n,
\end{equation}
\begin{equation}\label{I15}
J_{yx}=-D\nabla_yS_x-\Omega\tau N n,
\end{equation}
\begin{equation}\label{I16}
J_{yy}=-D\nabla_y-\frac{N\omega_s}{2}S_z,
\end{equation}
and
\begin{equation}\label{I17}
J_{yz}=-D\nabla_yS_z+\frac{N\omega_s}{2}S_y+
\frac{\omega_sN^2\tau}{2}\nabla_x n.
\end{equation}
Now we compare the Eqs.(\ref{I12})-(\ref{I17}) with those which would be
obtained, if we would simply read off the spin-current tensor from the 
divergence
of the diffusion equation, as suggested in Ref.[\onlinecite{Damker}].
Doing so, we note the following: while the result
for $J_{xy}$ and $J_{yx}$ is in both cases the same, the results for
the remaining components of the spin-current tensor are different
from each other (Note, that the fifth term on the lhs of Eq.(\ref{I10}) 
can also
be written as the divergence of a second rank tensor). There are two reasons 
for the differences. Firstly, the third
term in the Eqs.(\ref{I14}) and (\ref{I17}) has just the form of a curl
and thus can never be retrieved from the diffusion equation. Secondly, the 
terms linear in $N$ in the Eqs.(\ref{I12})-(\ref{I17}) differ from those
in the diffusion equation by a factor 1/2. This factor  can be traced back
to the following fact: the second, the third and the fourth term in 
Eq.(\ref{I10}) can be written
in the form $({\bm \Omega}\cdot{\bm S}-D\Delta{\bm S}-{\omega_s}
({\bm N}\times{\bm\nabla}))_l={D_i}_{lm}D{D_i}_{mk}S_k$, where
\begin{equation}\label{I18}
{D_i}_{lm}=\delta_{lm}\nabla_i+\frac{2Nm}{\hbar}\epsilon_{lkm}\epsilon_{kiz}
\end{equation}
is the covariant derivative, the image of the velocity
operator in the diffusion formalism. Thus, the spin diffusion 
equation (\ref{I10}) has a similar structure as the time-dependent
Ginzburg-Landau equation. In Eq.(\ref{I10}) the magnetic field is
replaced by the second term on the rhs of Eq.(\ref{I18}). Collecting
all terms, which are proportional to the derivative, means counting the
connection form twice, a fact which is known to be incorrect from the
Ginzburg-Landau equation. 

The Eqs.(\ref{I12})-(\ref{I17}) differ from those of 
Ref.[\onlinecite{Halperin}] in the second term on the rhs of the 
Eqs.(\ref{I13}) and (\ref{I15}). 
This term is absent in Ref.[\onlinecite{Halperin}]. However, it has important
consequences: since the particle density does not vanish at the 
boundary the magnetization can not be zero there. To investigate 
this point
further we now focus on the situation in the half plane $y>0$. Doing so, we 
ignore the impact of the spin-orbit interaction on the particle density\cite{remark}. First we focus on the situation in equilibrium. In this case 
$\nabla_x=0$, $d/dt=0$ and the particle density $n$ is equal to a constant 
$n_0$. Solving the spin-diffusion equation (\ref{I10}) subjected to the
boundary condition $J_{yk}|_{y=0}=0$ yields $S_y=S_z=0$ and
\begin{equation}\label{I19}
S_x=\tau N n_0\sqrt{\frac{\Omega}{D}}\exp(-\sqrt{\frac{\Omega}{D}}y).
\end{equation}
Thus, $S_x$ is non-vanishing at the boundary. If instead of half plane we 
would have chosen a strip of finite width $ 2 b$ we would have obtained
\begin{equation}\label{I19a}
S_x=-\tau N n_0\sqrt{\frac{\Omega}{D}}\frac{\sinh(\sqrt{\frac{\Omega}{D}}y)}
{\cosh(\sqrt{\frac{\Omega}{D}}b)}
\end{equation}
instead. 

Now we bias the system. To this end we use the method of Ref.
[\onlinecite{Halperin}], that is instead of switching on a true electric field
we introduce a concentration gradient. In this case the stationary solution to 
the particle diffusion equation is given by $n(x)=n_0+n_1x$. The presence 
of $n_1$ leads to a spin-accumulation, as discussed in 
Ref.[\onlinecite{Halperin}]. Due to $n_1$ we obtain $S_x=S_z=0$ and $S_y=
S_{bulk}=-n_1\tau N$ in the bulk. Despite this fact, the spin-Hall current 
vanishes in the absence of a boundary, since 
the second term on the rhs of Eq.(\ref{I17}) cancels the third. 
To investigate the situation in a half-plane
we impose again the boundary
condition $J_{yk}|_{y=0}=0$. In this case, we obtain
\begin{eqnarray}\label{I21}
S_z&=&-S_{bulk}\mbox{sgn}N\nonumber\\
& &\hspace{-1.5em}\mbox{Re}(
\frac{(1+\lambda^{*2})(1-\lambda^2)(\lambda^*-1)}{\lambda^*-\lambda}
\exp(-\lambda\sqrt{\frac{\Omega}{D}}y)).
\end{eqnarray}
Here  $\lambda=\sqrt{-1/2+i\sqrt{7/4}}$, and $^*$ is complex conjugation.
The result for $S_x$ agrees with Eq.(\ref{I19}) up to  a replacement 
$n_0\to n(x)$. If instead 
of the
half
plane we consider a strip of finite width we obtain again a result which
is antisymmetric with respect to $y$.

Here, however, the question arises what the source of the spin-Hall effect is,
if the spin-Hall current vanishes in the bulk. The answer is, that
the spin-Hall effect is not driven by the spin-Hall current, but by the
equilibrium magnetization, thus finally by $J_{yx}$. Once the 
magnetization exists the field induced rotation (the fourth term in
Eq.(\ref{I10})) turns the magnetization into the $z$-direction if the
system is biased.

The existence of the equilibrium magnetization, however, can not be
reconciled with time reversal invariance. The only way to restore
time reversal symmetry is to assume that the spin-current
density has a discontinuity at the sample boundary. The discontinuity
results from the coupling of the particle spin to the particle momentum,
which is provided by the electron spectrum.  Owing to the coupling  the spins
of the particles running toward the boundary are mainly aligned in positive
$x$-direction. The particles, which are reflected at the sample boundary
have their spins aligned in the opposite direction. Thus, the spins flip
when the particles are scattered elastically at the boundary. 
This fact has to be taken into account in the interpretation of the 
spin-current tensor (\ref{I1}). Whereas the electric current across the 
boundary
is a measure for the difference between the number of particles running
toward the boundary and the number of particles reflected at the
boundary the spin current tensor (\ref{I1}) is actually a measure for
the sum of both, for the number of particles running toward the boundary
and the number of particles reflected at the boundary. Therefore, the
spin-current tensor (\ref{I1}) does not characterize  a flux of spins across
the boundary  in the conventional sense in the Rashba model and jumps at
the sample boundary. 
The correct boundary condition is ${\bm S}|_{boundary}=0$ in the present
situation. In this case there is neither an equilibrium magnetization nor
a spin-Hall effect in a dirty Rashba semiconductor.

Clearly, at this point the question arises why there should be a 
non-vanishing
spin-Hall effect in an electric field in the same model in the clean limit, 
in which the energy
level splitting due to the Rashba interaction is large compared to the 
disorder energy. The electric field
itself does not affect the time reversal invariance. The  boundary 
condition, 
however, does in this case. Despite this fact energy conservation 
requires that the spins have flipped when the particles have been
reflected elastically at the sample boundary, at least if the particles
are separated from the boundary by a distance of the order of the mean
free path. Consequently, we expect that the spin accumulation  is
confined  to a  thin layer, which has  at most a width of the order of 
the mean free path. 

In summary, we have shown that the assumption that the spin-current
tensor is continuous at the sample boundary leads to results, which
are in contradiction to time reversal symmetry. The only way
to reconcile our results with  time reversal symmetry is to assume,
that the spin-current tensor has a jump at the sample boundary. 
The existence of the jump can be traced back to the coupling between 
spin and 
charge. We would like to stress that our results are specific to
systems with Rashba spin-orbit interaction and do not apply to systems,
in which the spin-orbit interaction is produced by the scattering
at impurities. The latter differ from the systems discussed here in that
there is no coupling between spin and momentum in the spectrum. 

. 

\begin{thebibliography}{}
\bibitem{remark} A consideration of the impact of the spin-orbit interaction 
on the particle density does not effect any of our arguments. It leads to a
weak accumulation of charge carriers at the boundary, which does not affect
the final result.
\end{thebibliography}
\end{document}